\documentclass{article}

\usepackage[english]{babel}
\usepackage[table]{xcolor}
\usepackage{array}
\usepackage[letterpaper,top=2cm,bottom=2cm,left=3cm,right=3cm,marginparwidth=1.75cm]{geometry}

\usepackage{amsmath}
\usepackage{multirow}
\usepackage{graphicx}
\usepackage[colorlinks=true, allcolors=blue]{hyperref}

\usepackage[utf8]{inputenc} 
\usepackage[T1]{fontenc}    
\usepackage{hyperref}       
\usepackage{url}            
\usepackage{graphicx}
\usepackage{natbib}
\usepackage{doi}

\title{Implementing Automated Data Validation for Canadian Political Datasets}
\author{Lindsay Katz and Callandra Moore}

\begin{document}
\maketitle

\begin{abstract}
This paper describes a series of automated data validation tests for datasets detailing charity financial information, political donations, and government lobbying in Canada. We motivate and document a series of 200 tests that check the validity, internal consistency, and external consistency of these datasets. We present preliminary findings after application of these tests to the political donations ($\approx10.1$ million observations) and lobbying ($\approx711,200$ observations) datasets, and to a sample of $\approx380,880$ observations from the charities datasets. We conclude with areas for future work and lessons learnt for others looking to implement automated data validation in their own workflows.
\end{abstract}

\section{Purpose}

The Investigative Journalism Foundation (IJF) has collated and actively maintains eight public interest databases relating to political donations, charity financial information, and government lobbying in Canada. These data are publicly available by the IJF in a form that is clean, interpretable, and can be queried and explored by users with ease. However, there is great variation in the accessibility, completeness, and cleanliness of the raw data sources upon which these databases are built, both across regions and over time. This has necessitated a complex data pipeline built by \href{http://www.theijf.org/databases}{the IJF} which routinely and programmatically updates each database while maintaining data cleanliness and standardization. This data pipeline executes a number of processing steps through which each piece of data must pass to reach its final form.

Automated data testing is a valuable tool for verifying that data are meeting certain standards or expectations held by the user, while simultaneously uncovering inconsistencies or errors within the data \citep{alexander2023telling}. This is especially beneficial for complex collated databases such as the IJF's, which integrate data from multiple origins across time. Moreover, the construction of all datasets involve fundamental assumptions and programmatic decisions which inform downstream analysis and use. To automate data validation for the IJF, we have developed a bespoke suite of automated tests spanning each of the eight IJF databases using Python’s \href{https://greatexpectations.io/}{Great Expectations (GX) library}. This means of data quality testing facilitates trust and transparency in the data being shared \citep{alexander2023telling}, and consequently in the news and scholarly articles informed by these data. 

In this report, we begin with a review of the current literature and computational tools available pertaining to data quality assessment. We then provide a detailed description of our workflow, following \citet{alexander2023telling}. This is followed by a discussion of future work, both in the IJF's unique data testing efforts, and data validation more generally. We then close with a conclusion outlining the main learnings from this work.

\section{Literature Review}
Concerns surrounding the transparency and replicability of published research have gained prominence in recent years, inspiring greater awareness and discussion of the need for reproducibility to be incorporated into scientific workflows \citep{vilhuber2022teaching, alexander2023telling, gelman2016}. The issue is highlighted by articles which attempt, and in many cases, fail, to reproduce various published research findings across disciplines \citep{vilhuber2022teaching, trisovic2022large}. Such work has shone light on the need for a transformation of the standards set for published research across disciplines, particularly reproducible workflows and accessible data and code. Data validation is a necessary tool for this. Chapter 3 of Telling Stories With Data \citep{alexander2023telling_full} is devoted to reproducible, well-documented workflows, and emphasizes that openness of code and data, especially detailing modifications to the original unedited data, are crucial components of reproducibility. Without such a transformation, researchers and journals may continue to publish works which are not replicable, in turn perpetuating public distrust of scientific research, and the publication of misleading conclusions.

\citep{alexander2023telling} also provides a detailed framework for writing a suite of data tests to improve the quality of one’s code by documenting the expectations they have of their data at particular points in the code \citep{alexander2023telling}. Specifically, focus is placed on testing for validity, internal consistency, and external consistency of the data. Validity refers to general correctness of variable classes and values (e.g., names do not contain numerals; numeric data is classified as such); internal consistency refers to coherence within the dataset (e.g., component columns summing to the total column); and external consistency relates to coherence of the data with relevant external sources \citep{alexander2023telling}. In providing such a framework, Alexander highlights the fundamental relationship between data transformation and data validation. Data transformation involves strategic decision making based on characteristics we would like the data to have, and data validation involves testing that those characteristics hold true in the data at large. 

Taking a more domain-specific focus, \citet{kohane2021every} discuss concerns surrounding the quality and reproducibility of research studies based on electronic health record (EHR) data. The authors advocate for six considerations to assess the quality of these studies, broadly pertaining to how complete, accurate, transparent, and comprehensive the data and analyses are \citep{kohane2021every}. Additionally, \citet{kahn2016harmonized} present a harmonized framework for EHR data quality assessment to encourage users to comprehensively evaluate the fitness of the data to their specific research goals. This framework includes data validation, emphasizing the importance of an alignment between characteristics of the data, and “relevant external benchmarks” \citep{kahn2016harmonized}. \citet{lee2017framework} implement the framework developed by \citet{kahn2016harmonized} within the heart failure domain, illustrating the importance of domain knowledge for developing a comprehensive, accurate set of tests for data quality \citep{lee2017framework}.  

Some computational tools have also been developed specifically for machine learning projects \citep{breck2019validation, hynes2017data}. \citet{breck2019validation} present an anomaly detection data validation system for data used in machine learning pipelines, deployed as part of TFX at Google. The authors emphasize the downstream effect that one data error can have on machine learning infrastructure, and the importance of catching assumptions made in the data wrangling process early on \citep{breck2019validation}. \citet{hynes2017data} also present a data validation framework for machine learning datasets, called Data Linter. The authors acknowledge that error detection in machine learning data is a time-consuming, error-prone, and iterative process, and present a tool which analyzes the data and offers variable transformation recommendations based on the specific model that will be trained \citep{hynes2017data}. These works illustrate the ways in which assumptions made about the data can get lost in the data science workflow, and the importance of checking and documenting them to avoid misleading or inaccurate conclusions.

In addition to the domain-specific frameworks for data validation, there also exist a number of more general-purpose computational tools for data testing. \citet{alexander2023telling_full} provides information and example code on how to use a number of libraries and functions for code testing in the R programming language \citep{r2022}, including \verb|testthat|, \verb|pointblank|, and base R’s \verb|stopifnot()| function. Notably, \verb|pointblank| contains built-in test functions which allow users to test that certain characteristics of their data hold. Additionally, \citet{marino2022r} provide a comprehensive review of R packages for assessing data quality with applications using publicly available cohort study data. The authors compare each package based on characteristics such as output format, string functionality, and availability of a graphical user interface \citep{marino2022r}. Great Expectations is a tool for validating, documenting, and profiling data in the Python programming language. This tool is useful as in addition to providing built-in validation test functions, it also offers an Onboarding Data Assistant tool that profiles your data to create a suite of bespoke validation tests \citep{GXOnboarding}.

Evidently, much great work has been done in the realm of data validation to build cross-disciplinary frameworks to inform data quality tests and computational tools for the programmatic implementation of those tests.

\section{Workflow}

\subsection{Description of the raw data}

This subsection provides a brief overview of the raw, unedited data from which the IJF's databases were created. All eight datasets are continually updated by the IJF with new information as data are added to their source websites. 

\subsubsection{Charities}

The IJF's charities database is composed of three datasets: charity tax returns, charity staff compensation, and gifts received by charities. All of these data were sourced from the Canada Revenue Agency (CRA), covering data from 1990 to the present. The Income Tax Act (1985) legally requires registered charities in Canada to file an annual information return. As outlined on the IJF's methodology page for this database, “A complete information return includes form \href{https://www.canada.ca/en/revenue-agency/services/forms-publications/forms/t3010.html?ref=investigative-journalism-foundation.ghost.io}{T3010 Registered Charity Information Return}, a copy of the charity’s own financial statements, \href{https://www.canada.ca/en/revenue-agency/services/forms-publications/forms/t1235.html?ref=investigative-journalism-foundation.ghost.io}{Form T1235, Directors/Trustees and Like Officials Worksheet}, and if applicable, \href{https://www.canada.ca/en/revenue-agency/services/forms-publications/forms/t1236.html?ref=investigative-journalism-foundation.ghost.io}{Form T1236, Qualified Donees Worksheet / Amounts Provided to Other Organizations} and \href{https://www.canada.ca/en/revenue-agency/services/forms-publications/forms/t2081.html?ref=investigative-journalism-foundation.ghost.io}{Form T2081, Excess Corporate Holdings Worksheet for Private Foundations}” \citep{ijf_charities_methods}. 

\begin{figure}[h]
\centering
\includegraphics[width=0.6\textwidth]{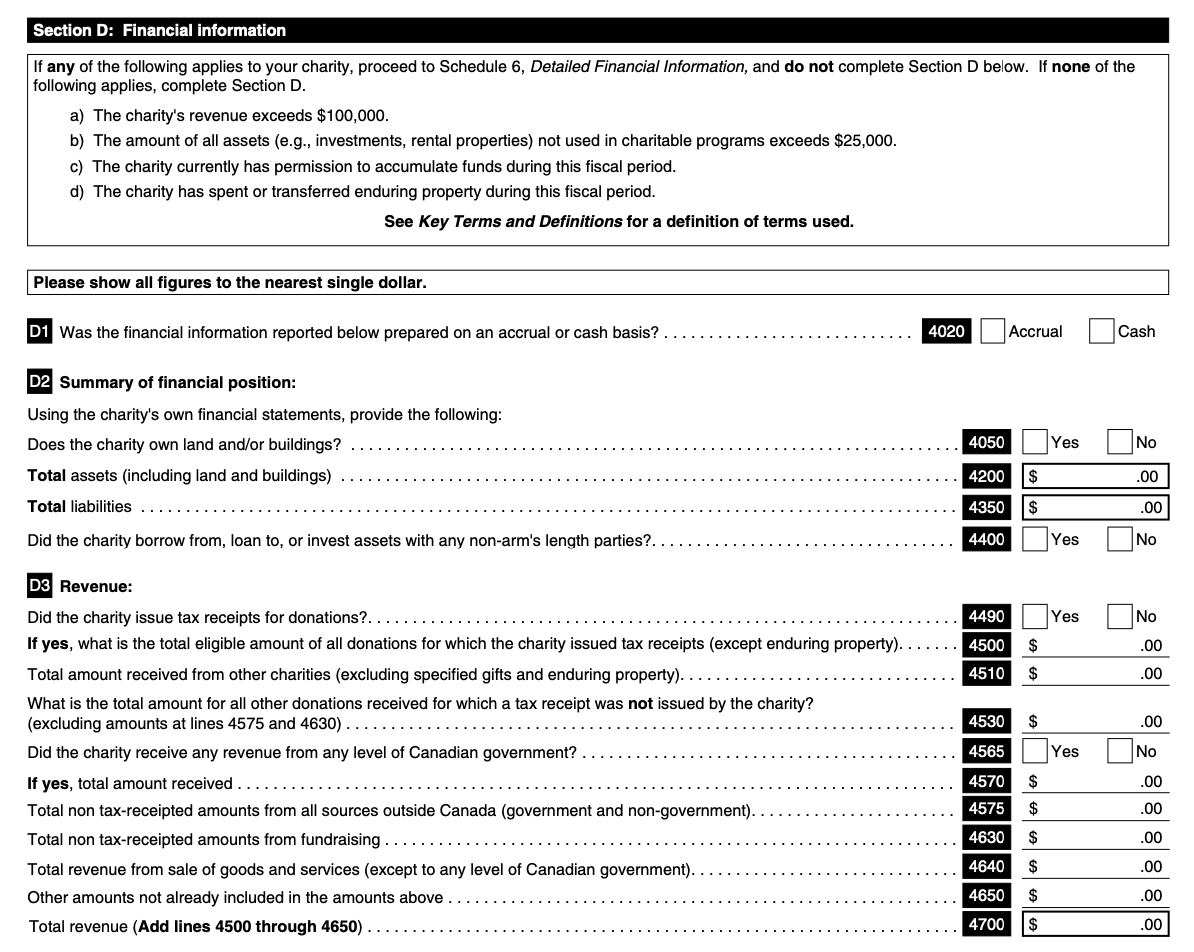}
\caption{\label{fig:image1}Part of the financial information section from the T3010 form from 2009.}
\end{figure}

The T3010 form is where all the data for the Tax Returns dataset is sourced. This form is composed of hundreds of distinct fields, called line items. These fields include a very intricate monetary breakdown of the charity’s assets, liabilities, revenues, expenditures, and gifts to the organization. Importantly, there have been a number of changes in the T3010 form since 1990, including the numbers and definitions affiliated with each line item. For instance, from 1990 to 1996, total liabilities was line item number 131, which changed to number 65 from 1997 to 2002, and then changed again to number 4350 from 2003 onward. A portion of the financial information section of the 2009 T3010 form, including line item 4350, can be seen in Figure \ref{fig:image1}.

\begin{figure}[h]
\centering
\includegraphics[width=0.6\textwidth]{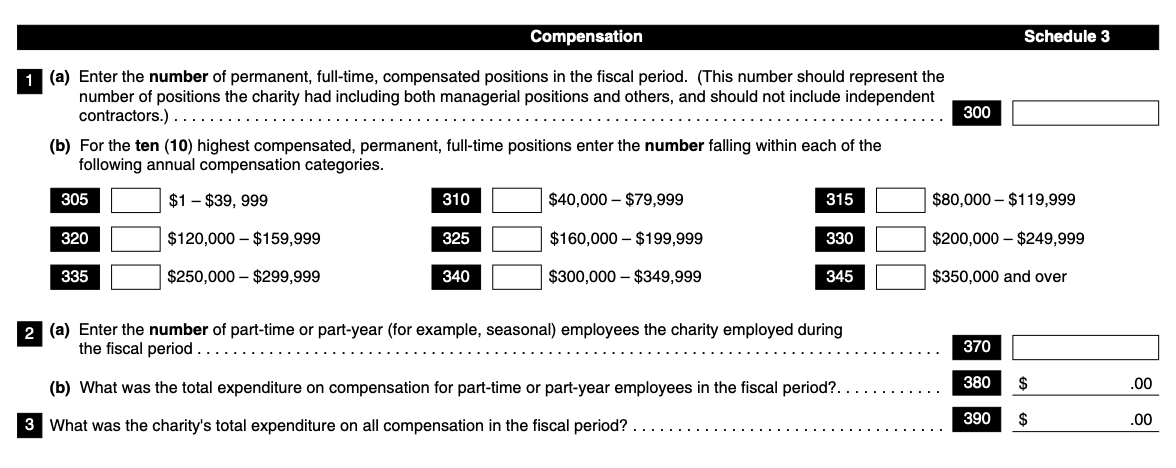}
\caption{\label{fig:image2}Compensation section from the T3010 form from 2009.}
\end{figure}

The charity staff compensation database contains data on the number of staff working at each charity, the total compensation for all positions, and the salary ranges for the highest paid employees. These data are sourced from the compensation section of the T3010 form, where the line item names and definitions have changed notably over time. In particular, the salary brackets defined by the CRA have encompassed different ranges over time. This portion of the T3010 form from 2009 is provided in Figure \ref{fig:image2}.

\begin{figure}[h]
\centering
\includegraphics[width=0.6\textwidth]{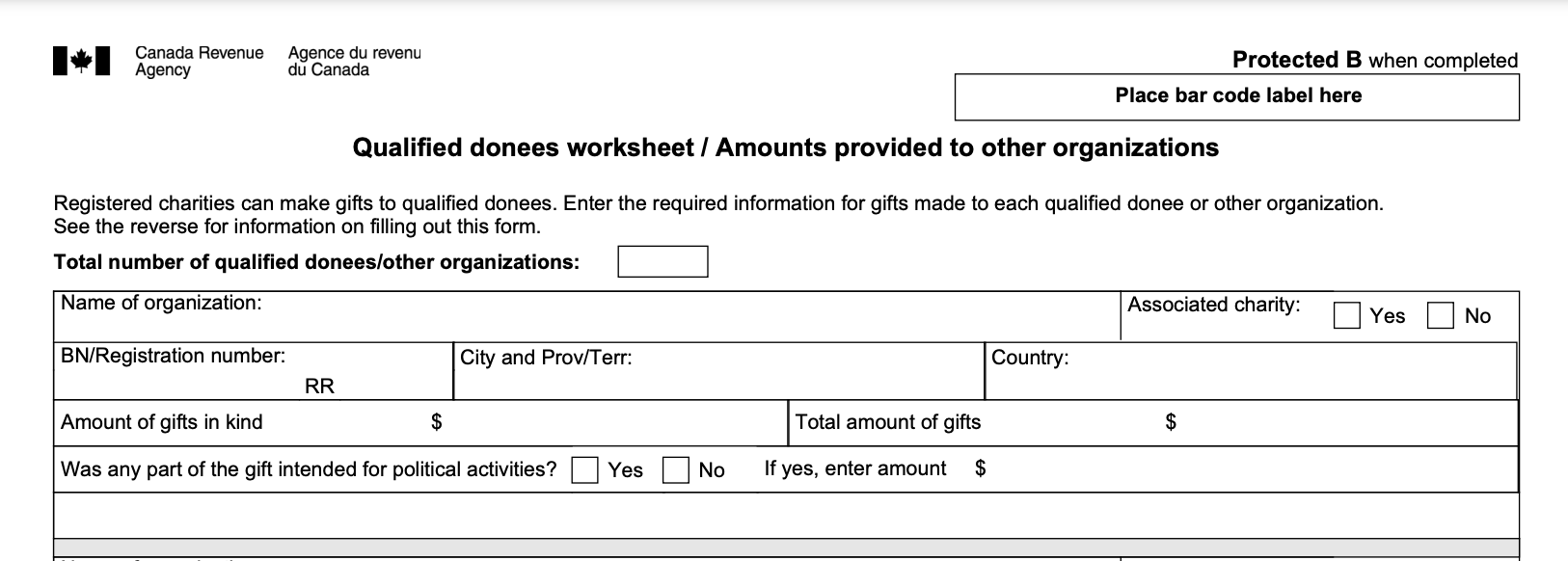}
\caption{\label{fig:image3}Portion of T1236 form from 2018.}
\end{figure}

Finally, the gifts received by charities data are sourced from the T1236 form which is filed by charities alongside the T3010 tax return form each year if they made donations to qualified donees in that fiscal year. An example of part of this form from 2018 is shown in Figure \ref{fig:image3}.

\subsubsection{Political Donations}

\begin{figure}[h]
\centering
\includegraphics[width=0.6\textwidth]{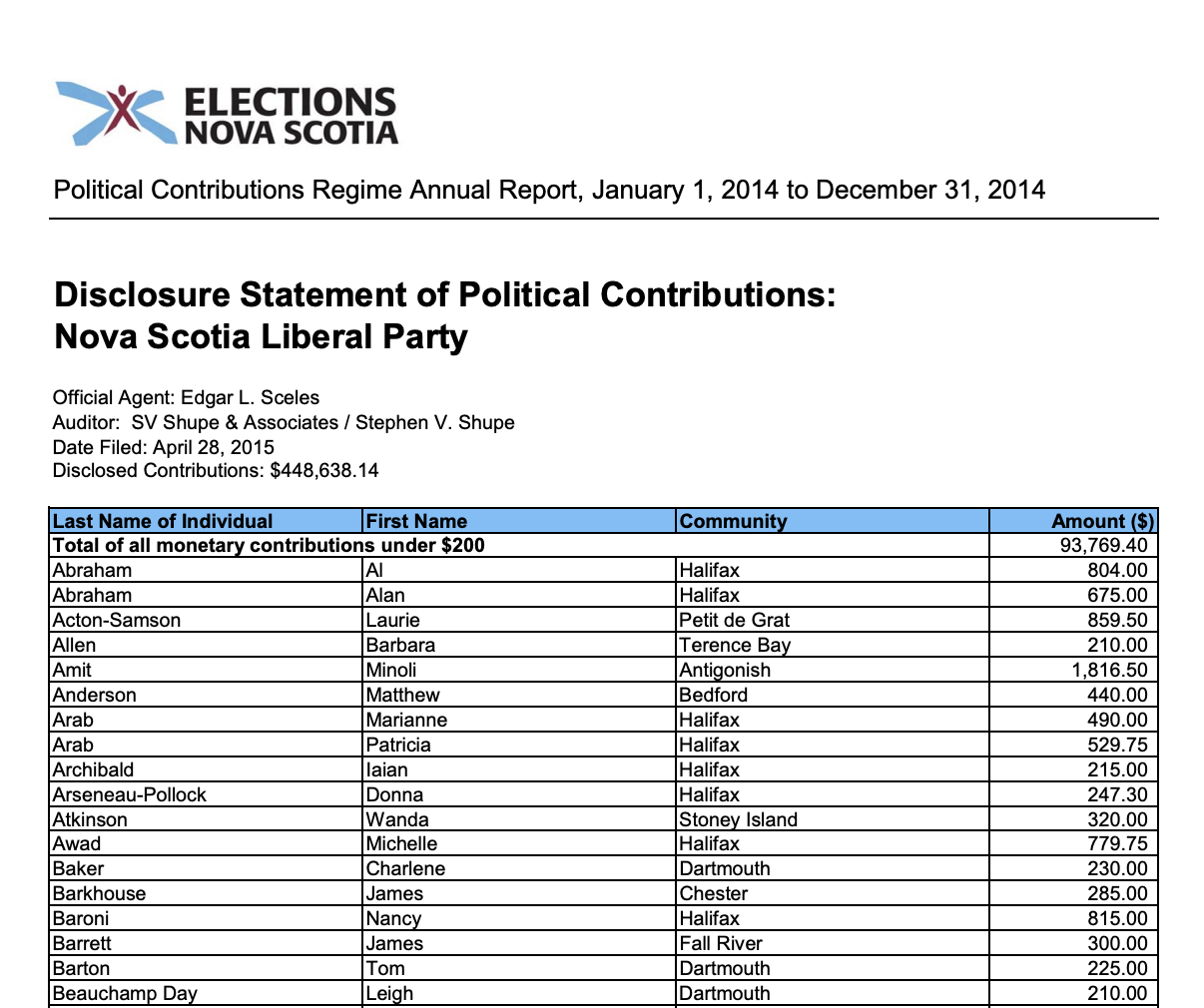}
\caption{\label{fig:image4}Portion of the 2014 political contributions PDF from Elections Nova Scotia.}
\end{figure}

The political donations database covers donations made federally, provincially, and territorially, with the earliest records from Elections Canada dating from 1993. Records of political donations are required by law to be submitted by political parties and/or candidates and are maintained and published by elections agencies \citep{ijf_donations_methods}. The frequency and scope of reporting required varies across jurisdictions, as does the type of recipients and donors that are allowed to receive and give political donations \citep{ijf_donations_methods}.  Maximum legal donation amounts vary across jurisdictions, and who is making the donation (e.g., an individual, corporation, or union). The IJF collected these donations data from elections agency websites, where files were stored as either a downloadable spreadsheet, PDF, or HTML form depending on jurisdiction and year. An example of the raw Nova Scotia donations data in PDF form is shown in Figure \ref{fig:image4}.

\subsubsection{Lobbying}

\begin{figure}[h]
\centering
\includegraphics[width=0.6\textwidth]{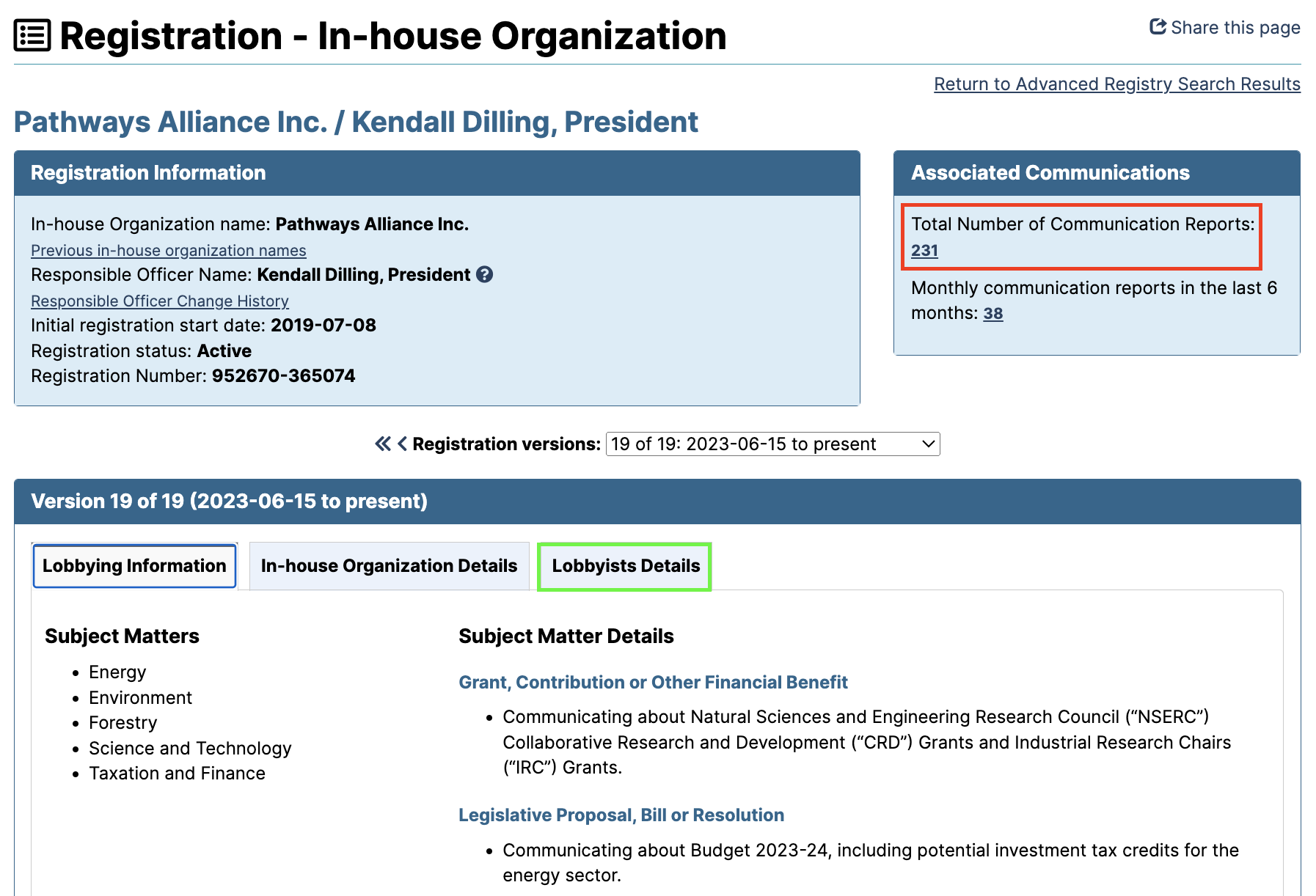}
\caption{\label{fig:image5}Screenshot from the Federal registry of lobbyists website.}
\end{figure}

The lobbying data is composed of four databases: lobbying registrations, government funding, lobbying communications, and revolving door (that is, lobbyists who formerly held government positions and have since transitioned into lobbying). In Canada, lobbyists must register with the lobbying registrar of all jurisdictions in which they are active, and disclose specific details on their activities \citep{ijf_lobbying_methods}. While there is regional variation in the information lobbyists are required to disclose, in general they are mandated to report for which organizations they are lobbying, the laws or subject matters that the lobbyists would like to discuss, and/or what money the lobbyists have or want to receive from the government \citep{ijf_lobbying_methods}. Figure \ref{fig:image5} provides an example of part of the webpage for a lobbying registration at the federal level.

The lobbying registrations and revolving door data were scraped from Federal, provincial, and Yukon lobbyist registries’ websites. Subject matter details and the list of government institutions being lobbied were collected for the registrations database, and details on the former public offices held by lobbyists were used to build the revolving door database. The designated public offices held data can be accessed through the “Lobbyists Details” tab outlined in green in Figure \ref{fig:image5}.

The government funding data were scraped from the Federal website and each province’s website. These data contain information on the amount of funding that the organization received by the government, broken down by source. Finally, lobbying communications data were scraped from the Federal and British Columbia registries (the only regions for which these data are available), and include details on communications between lobbyists and government officials that were disclosed by the lobbyists themselves. The red box in Figure \ref{fig:image5} illustrates where on the webpage the communications data can be accessed. These data “detail specific interactions between lobbyists and government officials”, making them a valuable supplement to the more general lobbying registrations data \citep{ijf_lobbying_methods}. Interactions with government officials can include email exchanges, meetings, and phone calls. 

\subsection{Initial data cleaning}

As illustrated in the previous section, the unedited, raw data used to build the IJF's eight databases came in many different forms, with structural variation across jurisdictions and over time. As such, the IJF performed some initial data cleaning where they deemed appropriate – that is, where cleaning would improve data usability, but not compromise data authenticity. 

Recall that the charity's tax return forms are composed of numerous line items whose numbers and definitions have changed over time. To keep track of this variation in the tax return forms, the IJF built spreadsheets which map every line item number to its correct definition and the years in which it was collected. This type of mapping spreadsheet is also known as a schema. The IJF selected about 250 line items of the over 600 available to include in their published dataset, which include main financial categories in the form and basic identification details about the charity \citep{ijf_charities_methods}. An example of part of the IJF's schema structure is shown in Table \ref{tab:schema}, where line item 4570 can be seen in Figure \ref{fig:image1}.

\begin{table}
\centering
\begin{tabular}{ | m{1.6cm} | m{1.6cm} | m{2cm} | m{4cm} | m{3cm} | }
\hline 
\rowcolor{lightgray} 
Start year & End year & IJF table & IJF column & Column name \\ [0.5ex] 
\hline
2003 & 2002 & liabilities & ch4300\_liabilities\_
accounts\_payable\_acrrued
\_liabilities & Accounts payable, accrued liabilities \\ 
\hline
1990 & 1996 & assets & ch123\_assets\_fixed\_other & Other fixed assets (land, buildings) \\
\hline
2003 & 2002 & revenue & ch4570\_revenue\_total\_
amount\_from\_govt & Total amount received from government \\
\hline
2003 & 2022 & expenditures & ch5050\_expenditures\_
gifts\_to\_qual\_donee & Total amount of gifts to qualified donees \\ 
\hline
1997 & 2002 & liabilities & ch65\_total\_liabilities & Total liabilities \\
\hline
\end{tabular}
\caption{\label{tab:schema}Part of the IJF's schema for the CRA tax return form line items from 1990 to 2022.}
\end{table}

Because all of these data are based on self-reported forms, and only about 1\% of charities are audited annually \citep{cra_audit}, they are prone to many human-made errors such as spelling mistakes, or incorrect dollar amounts recorded. The IJF performed data cleaning and standardization across the charity datasets where appropriate, to improve interpretability and consistency. In all three datasets, they deleted duplicated columns from the raw data and renamed some columns to make them more interpretable. The IJF converted fully capitalized text to lowercase when the text did not consist of proper nouns and converted names and cities to title case. For the tax returns data, they computed which columns add to the totals for each component of the tax return, such as which line items are sub-components of total liabilities. For the gifts received by charities data, the IJF removed rows where the donation amounts were distinctly wrong based on two possible characteristics. These rows are characterized by either a donation amount exactly equal to the charity’s unique nine-digit registration number (a likely error in data ingestion), or a donation amount greater than one billion dollars when the charity’s total revenues and assets summed to less than one million dollars \citep{ijf_charities_methods}.

Since many donations records were only available in static PDF form, optical character recognition (OCR) technology was necessary to convert them to comma-separated value (CSV) form. Converting documents with OCR can lead to a number of errors in the resulting CSV, such as a dollar sign (\$) being parsed as a letter S or number 5. The IJF performed extensive manual cleaning to correct these OCR errors wherever possible, checking the original PDF files throughout this process. The IJF also cleaned and standardized a number of columns for clarity. Dates were standardized to YYYY-MM-DD format, donor names in the form of “surname, firstname” were standardized to “firstname surname” format, and abbreviated party names were changed to full party names. Further, the IJF had to collate these data across all jurisdictions to create an amalgamated political donations database.

The self–reported nature of lobbying registrations, and the amount of free text present within the data, means that there is much variation in spelling of names and targets. In an effort to mitigate some of this variation, title casing was applied to government titles, and any unnecessary numbers were removed from the text. In the exact same manner as the donations data, dates and the structure of names were also standardized. Additionally, in the data cleaning process, the IJF discovered that the reporting forms in Quebec and New Brunswick require dollar amounts to be spelled out as text, while those forms in all other jurisdictions require dollar amounts to be written numerically \citep{ijf_lobbying_methods}. As such, the data for Quebec and New Brunwick had to be converted, both programmatically and manually, to numeric form by the IJF.

As illustrated by this overview of the IJF's initial data cleaning, the process of preparing and creating a dataset requires a number of choices to be made, many of which are informed by characteristics of the raw data only known to those who have access to it. As such, potential errors present in the raw data, in combination with the cleaning and standardization choices made by those involved in dataset construction — in this case the IJF — are crucial to consider when developing accurate, valuable expectations to set for the data.

\subsection{Test Development}

Our approach to developing the test suite can be broadly summarized within the framework introduced by \citep{alexander2023telling} grounded in tests for validity, internal consistency, and external consistency. To implement this framework in data tests, it was necessary to first examine the source forms of the data (e.g., charity tax return forms from 1993 to present which inform the schema), the desired format of the data (i.e., what is displayed publicly), and the methodology employed to create the latter from the former. This approach enables a stronger understanding of the context and structure of the data, what validity and internal consistency look like for the data, and what external data may be relevant for testing, ultimately leading to the development of more comprehensive tests.

\subsubsection{Validity}

Validity in the IJF's databases largely centers on expectations surrounding missingness. Missingness tests for validity are applied to variables created by the IJF that should be populated across all datasets, jurisdictions, and time for which missingness would imply an obvious error. In our test suite, we test the “rid” unique identifier variable and the “added'' datetime variable for missingness across all databases where they exist. An additional test for validity we developed checks date format. We developed a test to check that all values in the “date” column of the political donations dataset match the expected YYYY-MM-DD pattern. Another important component of database validity is checking variables classes. An incorrect variable class can lead to misleading results of statistical tests or models, and can be a detriment in data visualization (e.g., a discrete variable which is classified as continuous). The relational database management system used by the IJF checks for variable classes independently, however in general, variable class checks should be accounted for when developing a test suite.

\subsubsection{Internal Consistency}

There are a number of columns across all eight databases for which we can reasonably expect either a specific subset of the data, or all of the data, to not be missing. Such tests fall under the internal consistency approach as they focus on the scraped and processed data, for which expectations of missingness are more involved and may depend on other characteristics of the data. As mentioned, there is significant variation in the source data over time and across jurisdictions for all databases, meaning many expectations of missingness are only applicable to data from certain years and regions, in accordance with the IJF's schemas. As a result, these expectations depend on how the data exists internal to the final database, once it has been processed and collated. 

To develop missingness expectations for internal consistency, we first looked at all columns present in each dataset, and consulted with the IJF to create a list of those variables where these tests were appropriate. There are two types of tests for missing values we employ: those for data being missing and those for data not being missing. The latter type is applied to columns such as donation amount or lobbyist names, which have been scraped, cleaned, and collated by the IJF, and are expected to be present across all jurisdictions and time. This test type is also used to detect reporting completeness in charity tax returns, which is characterized by a charity reporting a total value for at least one of their expenditures, revenues, assets or liabilities for that fiscal period. We use the former to test for coherence between all line item columns in the charities tax returns database and the IJF's schema (see example Table \ref{tab:schema}) detailing in which years each line item was collected. We also do so for the salary range line item columns maintained in the charity staff compensation database. Since we know the timeframe in which each line item was present in the T3010 form, we expect rows in the data attributed to a fiscal period end date outside of that given line item’s timeframe in the CRA form, the data for that line item should be null. For instance, line item 65 seen in Table 1 represents total liabilities, and is coded in the IJF schema to have been recorded from 1997 to 2002. Therefore, if our test detects a non-null entry for that line item in a row with a fiscal period end date outside 1997 to 2002, there is an error in either the schema, or the parsed data.

Table \ref{tab:table2} provides a summary of all missingness-related tests we developed for internal consistency – that is, all tests developed for where data \textit{should} be missing, and tests developed for where data \textit{should not} be missing.

\begin{table}
\centering
\begin{tabular}{ | m{2.5cm} | m{2.5cm} | m{3cm} | m{4cm}| }
\hline 
\rowcolor{lightgray} 
Database & Expectation & Applicable data & Columns to be tested \\ [0.5ex] 
\hline
Donations & Expect no data to be missing & All rows & Amount, donor full name, region, political party, donation year, recipient, political entity, donation date\\ 
\hline
Lobbying Registrations & Expect no data to be missing & All rows & Registration number, org name, region, subject matters, targets, affiliates, categories \\
\hline
Lobbying Communications & Expect no data to be missing & All rows & Subject matters, name, lobbyist, targets \\ 
\hline
Government Funding & Expect no data to be missing & All rows & Region, entity, registration number, sum, source, financial end \\
\hline
\multirow{2}{2.5cm}{Charities Tax Returns} & Expect that tax returns do not have data for line items which were not present on the T3010 associated with the fiscal year & All rows conditional on the years specified in the IJF's schema for each given line item & All line item variables \\
\cline{2-4} & Expect that the value for at least one of total assets, expenditures, revenue, and liabilities is not missing. & All rows except for those attributed to the first return submitted by a charity upon charitable registration, or associated with a charity's status revocation. & All line items representing total assets, expenditures, revenue, and liabilities \\
\hline
Charity Staff Compensation & Expect that the compensation section of tax returns do not have data for line items which were not present on the T3010 associated with the fiscal year. & All rows conditional on the years specified in the IJF’s schema for each given line item. & All line items representing the number of staff paid in various salary ranges. \\
\hline
\end{tabular}
\caption{\label{tab:table2}Summary of missingness tests developed for internal consistency.}
\end{table}

\begin{figure}[h]
\centering
\includegraphics[width=0.5\textwidth]{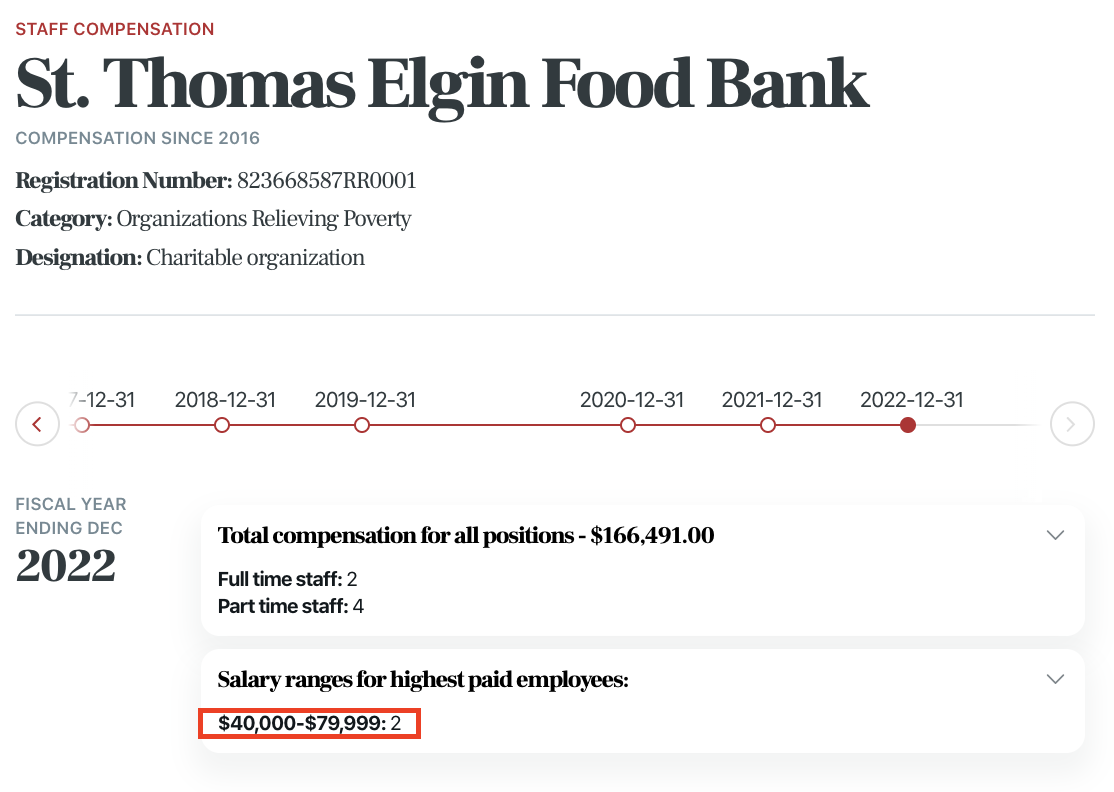}
\caption{\label{fig:image6}Staff compensation data for St. Thomas Elgin Food Bank in 2022.}
\end{figure}

\begin{figure}[h]
\centering
\includegraphics[width=0.5\textwidth]{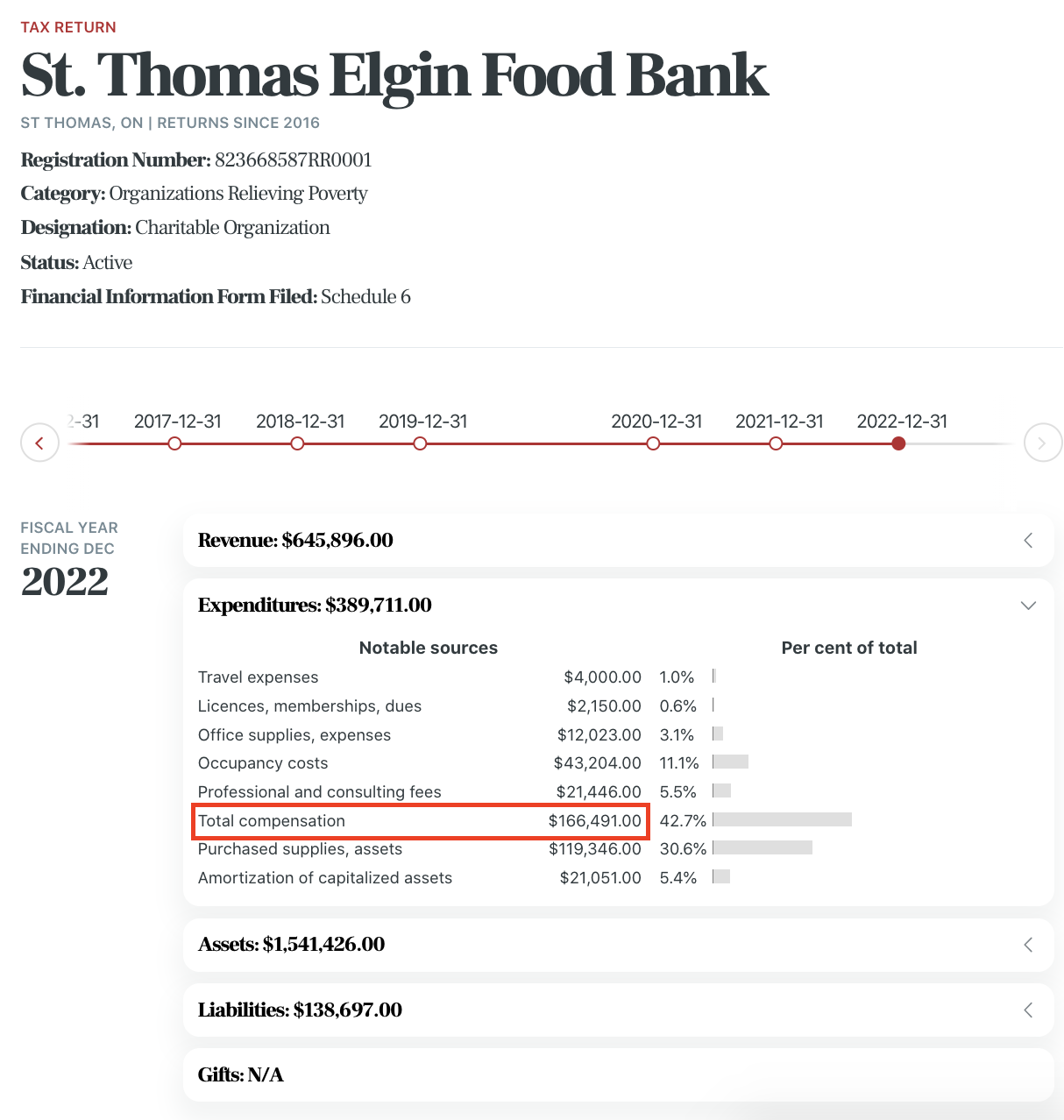}
\caption{\label{fig:image7}Expenditures portion of St. Thomas Elgin Food Bank’s 2022 tax return.}
\end{figure}

Another core characteristic of an internally consistent database is the sub-components of a value add up to the correct total. In the political donations data, in addition to the total `amount' variable, there are separate variables for the monetary and non-monetary contributions. Based on the availability of these variables, we developed a test that assesses whether the amount monetary and amount non-monetary sum to the total amount variable. We also checked for summation consistency across the charities databases. Knowing that there are line items which capture parts of a whole (e.g., lines 4490 to 4650 in Table \ref{fig:image1}), we set the expectation that the sum of gifts given by the organization in the gifts received by charities database is less than or equal to the ``total amount of gifts to qualified donees'' value in the expenditures portion of the tax return database. Additionally, we developed a test which checks for internal consistency between the charity’s compensation data from their tax return (Figure \ref{fig:image2}), and the charity’s reported compensation expenditures. In particular, we expect that the number of staff paid in each salary range multiplied by the lowest end of that range in the staff compensation data is less than to the ``Total compensation'' amount in the expenditures portion of the tax return data. An example of this is provided in Figures \ref{fig:image6} and \ref{fig:image7} below — we can see that total reported compensation from the highest paid employees is equal to $2\times40,000=80,000$, which is less than the total compensation amount of $166,491$, so our expectation is met.

\begin{figure}[h]
\centering
\includegraphics[width=0.9\textwidth]{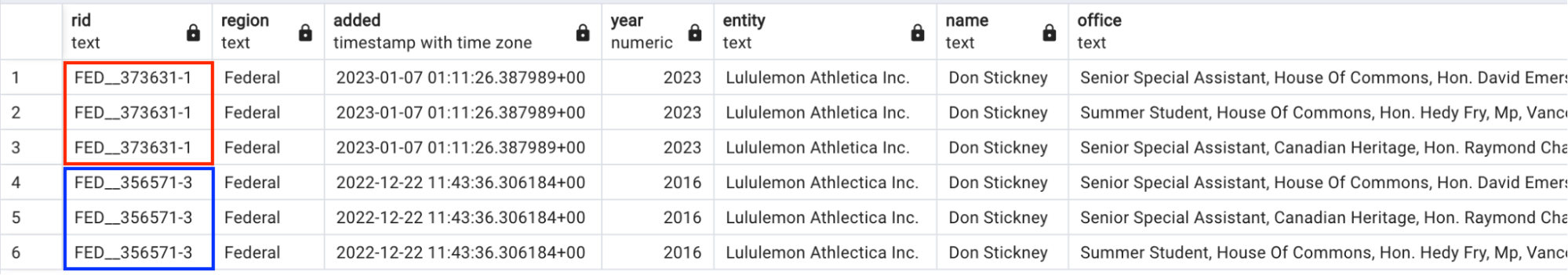}
\caption{\label{fig:image8}Screenshot from revolving door database.}
\end{figure}

\begin{figure}[h]
\centering
\includegraphics[width=0.9\textwidth]{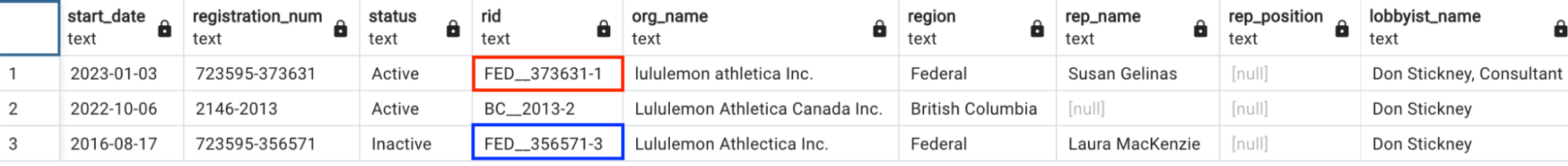}
\caption{\label{fig:image9}Screenshot from the lobbying registrations database.}
\end{figure}

In developing a test for internal consistency across the lobbying databases, we verified with the IJF team that we can reasonably expect that for each unique lobbying registration present in the revolving door database (based on record identification (RID)), there should be an entry with the same RID in the lobbying registrations database. As such, we added this expectation to our test suite. Figure \ref{fig:image8} is a screenshot from the revolving door data, and Figure \ref{fig:image9} is a screenshot from the lobbying registrations data. Notice the two distinct RID’s found in the revolving door database for Don Stickney at Lululemon, outlined in red and blue respectively, can be found in the registrations database outlined in the same colors. This is an example of what informed our expectation, and what exactly we are testing for with this expectation.

The final test for internal consistency we developed spans all databases for which there is a region variable (i.e., all lobbying databases and the political donations database). We test that all values in the `region' variable are equal to one of the official English names of the Canadian provinces and territories, with correct capitalization, such as ``Newfoundland and Labrador''. 

\subsubsection{External Consistency}

The final element of our test suite checks for external consistency. Tests of this nature require relevant external benchmarks against which we can check data values. We developed such a test for the political donations data. The IJF's \href{https://theijf.org/donations-methodology}{methodology page} summarizes the legal limits for political donations in each region according to political finance regulations as of November 2022. Using this summary, we intend to test all 2022 donations data against the legal limit for its jurisdiction. We do not screen earlier years’ data in this manner because summarizing the evolution of legal donation limits over time for each jurisdiction is a non-trivial research task. This is because legal donation limits evolve not only over time and region, but also type of donor (e.g., individual, corporate, etc.). Given the need for external data to develop these tests, they require additional research, and as such we plan to develop more tests of this type in future work. Details on this will follow.

\subsection{Implementation Process}

To implement our data tests programmatically, we employed Python’s Great Expectations (GX) library. GX provides a variety of pre-built expectation functions that are easy to implement, with a corresponding \href{https://legacy.docs.greatexpectations.io/en/latest/reference/glossary_of_expectations.html}{glossary} outlining what each function tests, its arguments, and its outputs. Importantly, many GX functions can be supplied with a `row\_condition’ argument, allowing the user to apply the function to only those rows which meet the specified condition.

Before running the test suite, we needed to first transform some of the data to be passed to its corresponding GX function. For the donations data, we had to remove all non-numeric characters (i.e., dollar signs and commas) from the amount column, and then convert that column to numeric. We also had to create an additional column in this dataframe equal to the absolute difference between the amount value and the sum of the amount monetary and amount non monetary values. While Great Expectations has a function that checks the sum of multiple columns, that function only has the capability to check those row-wise sums against a single value for all rows in the dataset, meaning we cannot compare column sums to a unique total value (in this case, the amount value) in each row. Performing data manipulation enabled us to perform the test using a different GX function without compromising on its design. For the charity data, we had to similarly remove dollar signs and commas from all line item variables and convert them to numeric. We also had to convert the fiscal period end column to datetime format. Due to the scale of the charities database, and the fact that it is maintained by the IJF in a number of distinct tables, evaluating expectations required extensive data manipulation including merging multiple dataframes on distinct charity registration numbers, transposing dataframes, and computing aggregates.

Having written the code to prepare the data as necessary, we implemented our tests on a random sample of about 10,000 rows of data based on ID number where available. In doing so, we found a large number of exceptions to many of these tests. For instance, we found that a large proportion of the fiscal period end date data in the government funding database were missing, and we found a number of rows in the political donations data where 2022 donations exceeded the legal limit for that year. Such findings prompted us to explore whether these exceptions indicate true errors in the data or are indicative of a need to adjust our expectations based on some characteristic of the data we had not been aware of. In many cases it was the latter. To determine this, we looked at the data which failed each test and performed exploratory analysis to detect patterns in these data. This process uncovered very informative and interesting trends. For instance, we found that for some columns with large proportions of missing values, those rows with missing data belonged almost entirely to a subset of regions and/or years. We also found that a large number of rows in the donations dataset with an amount value over the legal donation limit had the name ``Contributions Of \$200 Or Less/Contributions De 200 \$ Ou Moins’’.

Presenting these observations to the IJF team who are equipped with fundamental domain knowledge on the data, we were able to identify inaccuracies in our expectations, and adjust them accordingly. Expectations for two columns in which we set the expectation that no data should be missing were updated to account for different regional reporting requirements, recalling that these databases are an amalgamation of data across jurisdictions. For instance, the fiscal period end date variable in the government funding data is only collected in the Federal and Saskatchewan jurisdictions, meaning we can only truly expect there to be non-null values for that variable in those two jurisdictions. We adjusted the code for that test to reflect this. Additionally, we learned that donations with a donor name of ``Contributions Of \$200 Or Less/Contributions De 200 \$ Ou Moins’’ are aggregated donation values, meaning that the legal limit should not be applied to them.

After making these initial adjustments and re-running our test suite, we performed additional exploratory data analysis on tests that did not produce a 100\% success rate. Doing so allowed us to confirm that we had correctly accounted for characteristics in the data that we had previously excluded from our expectation code due to a lack of domain knowledge, and to check for any other trends present in the data that did not pass a given test. Having confirmed that there were no remaining patterns that implied inaccurate expectation conditions, we generalized our code by applying it to a larger sample of the data. We then iterated on our data expectations by testing larger samples of the data with each iteration and exploring the flagged data to identify interesting patterns. 

This process led us to detect additional characteristics of the donations data which improved the accuracy of our data tests. For instance, in subsequent iterations of the test for external consistency of 2022 donations amounts and the 2022 legal limits, we identified a number of exceptions where the donor name included ``Estate of’’. Upon further investigation, we learned that continuing contributions from a testamentary trust made before 2015 in the Federal jurisdiction have been subject to different legal limits \citep{furrow2015}, and those made before November 2017 in British Columbia were not subject to a limit at all \citep{carman2020}. Additional iterations uncovered other characteristics of the political donations data and legal regime unbeknown to us, requiring modification of our test code.  Further, many rows in the donations data had missing values for monetary amount and non-monetary amount. This disaggregation of amount type was only collected by Elections Canada after the year 2000. Based on this, we adjusted the expectation code to only run the test on post-2000 data. These and other discoveries highlight our lack of knowledge about political donation jurisprudence and record-keeping and the necessity of this knowledge for comprehensive data validation. Table \ref{tab:table3} provides a summary of the tests that required adjustment following this iterative implementation process, and describes what adjustments were made and why.

\begin{table}
\centering
\begin{tabular}{ | m{2cm} | m{6cm} | m{6cm}| }
\hline 
\rowcolor{lightgray} 
Database & Original Expectation & Condition added and explanation \\ [0.5ex] 
\hline
\multirow{3}{2cm}{Donations} & Expect donation dates to never be missing. & Region must be equal to one of Federal, Ontario, or British Columbia, as these are the only jurisdictions which collect this variable. \\
\cline{2-3} & Expect all donations data from 2022 to be less than or equal to the legal limit for that jurisdiction. & Donor full name must not contain ``Estate of'', ``Contributions of'' or ``Total Anonymous Contributions''. Estate contributions have distinct legal limits, and names which contain the latter two phrases are aggregates. Also, the political entity entry must not contain ``Leadership'', because the legal limits for donations differ for leadership contestants. \\
\cline{2-3} & Expect that for all the donations data, the maximum absolute difference between ``amount'' and the sum of ``amount monetary'' and ``amount non-monetary'' is 5. & The year must be greater than 2000, the jurisdiction must be Federal, and at least one of the ``amount monetary'' and ``amount non-monetary'' values must not be null. \\
\hline
Government Funding & Expect the financial end to never be missing. & Region must be equal to one of Federal or Saskatchewan, as these are the only jurisdictions which collect this variable. \\
\hline
\end{tabular}
\caption{\label{tab:table3}Overview of adjustments made to data tests following initial implementation.}
\end{table}

Evidently, the process of implementing our data validation test suite is iterative in nature, and required extensive fundamental domain knowledge to ensure the final tests were as accurate and informative as possible.

\section{Preliminary Findings}

At present, we have implemented all our expectations for the donations and lobbying databases, and most of our expectations for the charities databases. We are actively working to implement the final few expectations for charities in Python. Note that our preliminary findings are based on data queried from the development environment which may be slightly different from what is shown in production.

For the donations data, our expectations that for all rows, the values of donation amount, donor full name, political party, region, donation year, and recipient should not be null, was met with a 100\% success rate. The expectation that the political entity must not be null had a 97.76\% success rate in the data. Further, our test did not catch any exceptions to our expectation that where applicable, monetary and non-monetary donation amounts summed correctly to the reported total within a margin of error of $\pm5$ dollars. All donation date values matched the expected regular expression pattern based on the YYYY-MM-DD date format. Finally, for all regions except Federal, British Columbia, and Quebec, there were no donations in 2022 that exceeded the legal limit. Those three regions had 2, 1, and 11 exceptions, respectively. The two at the Federal level belong to individuals who were electoral candidates at the time meaning they can legally donate beyond \$1675. And the one exception for British Columbia appears to be a duplicate of another entry belonging to an estate donation. The exceptions for Quebec require additional investigation.

For lobbying registrations, the only tests which did not have perfect success rates were those where we tested that all registration numbers, organization names, and regions must not be null. These had 99.66\%, 99.44\%, and 99.45\% success rates, respectively. Across the other three lobbying databases, there were no missing region entries. For government funding, only one test caught exceptions, and that was due to null source entries we did not expect to be missing (98.91\% success rate). Two lobbying communications data tests were not perfect – we found two null target entries, and 25 null subject matter entries in the data. Finally, the expectation we set that for each unique lobbyist per organization in the revolving door database, there should be an entry with their name in the lobbying registrations database, had a 99.58\% success rate.

Across all donations and lobbying databases, we tested the expectation that all entries for the region variable must be the official English name of one of the Canadian provinces and territories. The only database which failed this test was lobbying communications, where we found 644 rows with a region listed as “Bc\_Reports”. This exemplifies the value of testing internal consistency, especially with a database as large as the IJF's. Having detected this inconsistency in region name, the IJF can now defer to the raw data, identify where this inconsistency originates, and modify their data cleaning pipeline accordingly such that newly scraped data which have this region name are subject to appropriate standardization.

Because of the scale of the charities databases, we took a random sample of these data by randomly selecting 20,000 registration numbers and querying all tax returns associated with those registration numbers. For this random sample of charities data, our tests detected 61 line item variables where there were non-null values in line items that were not collected on the associated year’s reporting forms (according to the IJF's schema). These all warrant further investigation, to detect whether the exceptions stem from an issue in the raw data (e.g., the charity completing an out-of-date form), or the IJF's schema. It should be noted that CRA data includes a number of records from before 1990 which is beyond the scope of the IJF’s database. For this reason, the IJF’s schema does not account for pre-1990 records, meaning our findings based on this sample and the IJF’s schema may have inflated error rates. Our expectation that the sum of gifts given by the organization in the gifts received by charities database is less than or equal to the reported total amount of gifts to qualified donees expenditures line-item value had a $\approx95.0\%$ success rate. This may in part be attributed to incomplete T1236 forms being filed. The expectation that the reported total compensation amount in expenditures is greater than the calculated lower bound was run separately from 2002 onwards. Our sample of data from 2003 to 2008 and 2009 to 2022 produced an $\approx97.8\%$ and $\approx98.5\%$ success rate, respectively. Finally, we test that the value of at least one of total assets, expenditures, revenue or liabilities is not missing unless it is a charity's first return filed. This test had an $\approx 97.3\%$ success rate.

Evidently, many of the expectations we set for the data were met when tested with code. While some of these expectations may seem simple, testing them programmatically and reporting the results as we have enables users to have a clearer image of the data with which they are working. Further, the tests which did not have a 100\% success rate have prompted us to dig deeper into the data to decide whether to adjust our expectations, modify the IJF's methodology, or flag inconsistencies or errors inherent to the raw data.

\section{Future Work}

While this project is still a work in progress, there are some interesting avenues of future work we would like to pursue. For the political donations data, we currently check data from 2022 against the legal limits for that year because we do not have a summary of the evolution of legal donation limits over time. In future, we would like to create this schema and use it to check donation amounts both before and after 2022 against the legal limits for the corresponding year. Creating this schema will be time-consuming, as we will need to check the legal limits for each year individually across all regions and account for any differences based on who is donating (e.g., individuals or candidates) and when they are making the donation (e.g., different electoral events).

\begin{figure}[h]
\centering
\includegraphics[width=0.9\textwidth]{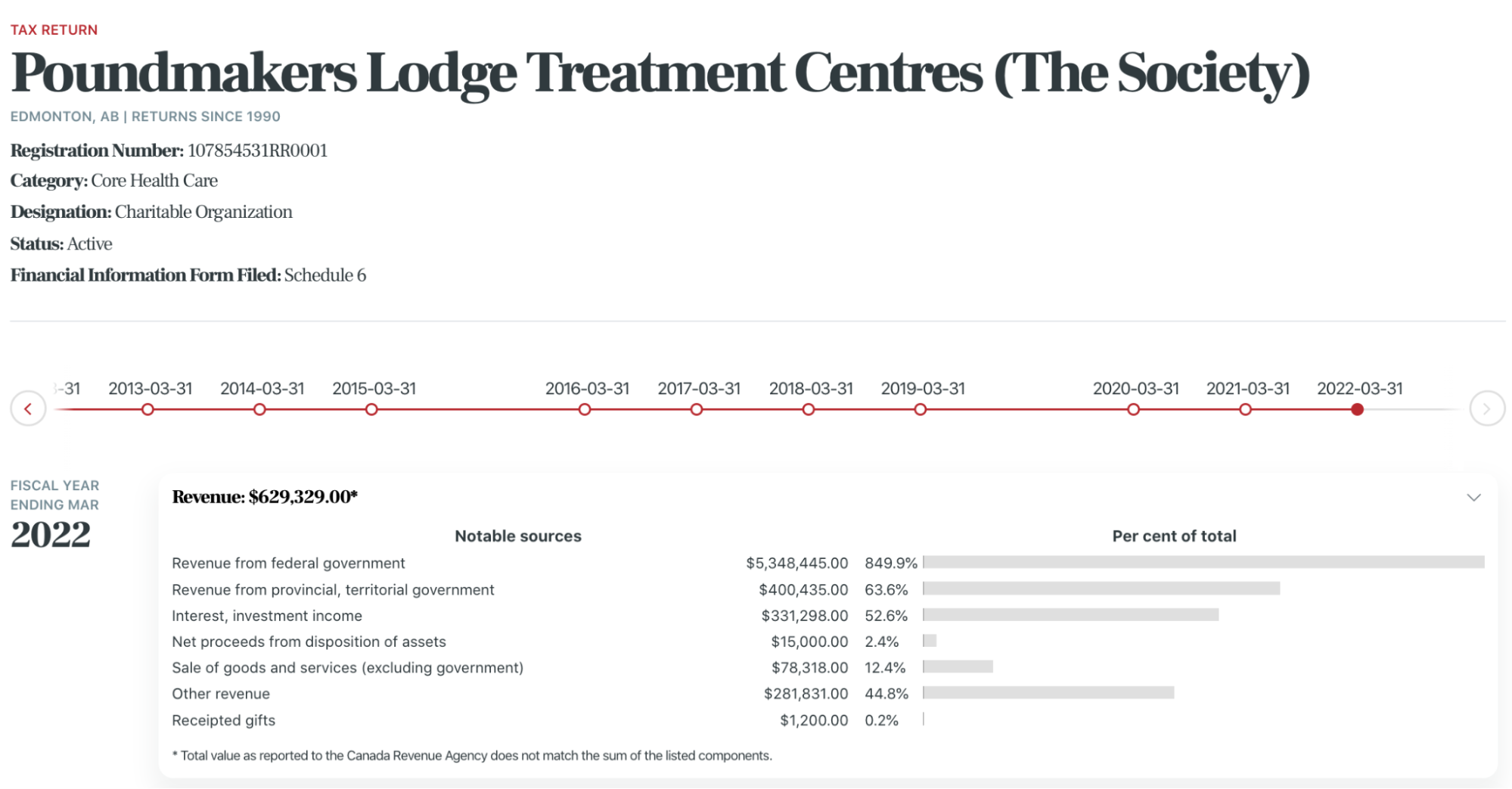}
\caption{\label{fig:image10}Example of IJF charities tax return data where sub-parts do not add up to the reported total (see asterisk).}
\end{figure}

Another avenue of future work would be to develop more detailed, comprehensive tests across the charities databases. We would like to set additional expectations that facilitate validation across line items in charities’ tax returns (e.g., expenditures, total gifts given to qualified donees, etc.). It would be valuable to implement automated checks to test that the sum of non-null line items add up to the total value reported using Great Expectations. The IJF has checked this for all the tax returns up to February 2023 themselves, an example of which can be seen in Figure \ref{fig:image10}, signified by the asterisk next to the total revenue. However, these tests were completed independently of the data cleaning pipeline. Running and deploying these tests with a tool like Great Expectations would allow for newly ingested data to be checked automatically as well. This is a particularly challenging task because for a high proportion of returns where the total is inconsistent with the addends, the tax return itself reports inconsistent or incorrect values and the error in summation is not an error on the part of the IJF's scraping, ingesting, or cleaning process. This problem is thus one of external rather than internal consistency. However, running this validation test is impractical at scale for a dataset of this magnitude. Also, the work of auditing tax returns is one of the CRA and not the IJF. 

It is also of interest to develop more free text-focused expectations, which would allow us to detect more inconsistencies in the data, especially within and across the lobbying databases which contain a large amount of text data. For instance, the IJF methodology pages for the charities and lobbying databases mention changing names written in the form  ``last name, first name’’ to the form of ``first name last name’’. Developing an automated data test to check for this pattern would enable the IJF to catch any rows of the data that they missed in the data standardization process.

Finally, we would like to develop an algorithm for checking duplicate rows in the data that contain only one minor difference, such as extra whitespace between two words. This is beyond the pre-defined functions made available in Great Expectations, and would require additional programming work, especially since we would ideally develop this algorithm for each database published by the IJF.

For data validation more generally, an interesting future endeavor would be to explore the capacity for large language models (LLMs) such as GPT-4 to produce a suite of data validation tests for a dataset via prompt-based in-context learning. While valuable, existing tools for developing automated data testing such as Great Expectations are inadequate on their own for producing a comprehensive suite of data tests which are as accurate as possible for the data at hand. This is because they are limited to a set of predefined test functions which do not incorporate domain knowledge, and as exemplified in the need to impose row conditions throughout our test suite, domain knowledge is a fundamental prerequisite for accurately understanding and assessing data. Further, developing a suite of tests can be quite time consuming and difficult, particularly for individuals who do not prioritize validation in their database construction process or who were not involved in the database construction process. As such, exploring the quality and breadth of data tests produced by LLMs in comparison to those created by an individual, such as those developed in this work, or other simple automated data validation suite may encourage others to prioritize data validation in their data pipelines, especially due to the decreased mental effort associated with LLM outsourcing.

\section{Conclusion}

Data validation is an important component of all workflows which use data for producing reproducible, transparent, and high-quality work. As illustrated in this project, not only does implementing automated data validation allow the IJF to check the quality, validity and consistency of their data, but it also facilitates transparency in outlining the true methodological assumptions that underpin each database. 

The process of developing a data test suite for the IJF presented a number of valuable lessons: 

\textbf{Understand your data backwards and forwards}. Familiarizing oneself with both the databases presented to the public and those used internally is crucial when beginning to build an expectation suite. Doing so allows for a balance of focus and understanding between the raw and amalgamated data, which leads to the development of more thorough data expectations. 

\textbf{Don’t be afraid to wrangle}. Data wrangling is sometimes necessary to run tests in the way that you desire — more so when implementing those tests with a tool such as Great Expectations, that has predefined functionality. Data tests should not be modified or compromised to fit the limited test functions available. Users should harness the powerful tool of data wrangling to manipulate their data in such a way that it fits the format necessary for validation. 

\textbf{Iterate}. This work has exemplified that the process of implementing data validation is necessarily iterative, and is continually being updated by failed tests and new knowledge. This is simplest when breaking the data into manageable chunks and gradually increasing sample size. Exploring the flagged data with each iteration presents an opportunity to identify important trends and characteristics inherent to the data, which can inform test development.

\textbf{Expertise is key}. Arguably the most important lesson learned, however, is that relying on the data alone is insufficient for producing a comprehensive suite of tests. Domain knowledge is a crucial component for developing accurate data tests and interpreting the results of those tests. In our case, this involved collaborative efforts with the IJF to acquire knowledge of political donations regulations, historical tax return forms, and regional differences in reporting requirements across all the databases. The danger of a user or data scientist making assumptions about the data based on personal expectations informed only by the data is that they do not always hold true in practice, and could result in misleading conclusions. 

Though the data validation process is by its nature never completed, our work offers a fundamental basis for testing that core beliefs about the data hold true at scale. Further, this work has illustrated the extent of resources necessary to build a test suite that is both valuable and accurate. Despite this difficulty, data validation work is vital to the reproducibility, transparency, and quality data analysis workflows across research domains from machine learning to political science to neuroscience. Our work here can serve as a framework to other research projects undergoing similar challenges. The complexity and importance of validation in all data-focused workflows illustrates the need for developments in the realm of making data validation easier and more accessible to implement for individuals of all backgrounds.

\bibliographystyle{unsrtnat}
\bibliography{gx_refs}

\end{document}